# John Bell's Varying Interpretations of Quantum Mechanics

– Memories and comments –

H. Dieter Zeh (www.zeh-hd.de) - arxiv:1402.5498v8

**Abstract:** Various interpretations of quantum mechanics, favored (or neglected) by John Bell in the context of his nonlocality theorem, are compared and discussed.

## 1. Varenna 1970

I met John Bell for the first time at the Varenna conference of 1970 (d'Espagnat, 1971). I had been invited on the suggestion of Eugene P. Wigner, who had already helped me to publish my first paper on the concept that was later called decoherence – to appear in the first issue of *Foundations of Physics* a few months after the conference (Zeh, 1970). This concept arose from my conviction, based on many applications of quantum mechanics to composite systems under various conditions, that Schrödinger's wave function in configuration space, or more generally the superposition principle, is universally valid and applicable. In particular, stable narrow wave packets can represent classical configurations, while their superpositions may define *new individual* properties – such as "momentum" defined as a plane wave superposition of different positions. Superpositions of macroscopically different properties, on the other hand, are regularly irreversibly "dislocalized" (distributed over many degrees of freedom) by means of interactions described by the Schrödinger equation. The corresponding disappearance of certain *local* superpositions ("decoherence") seems to explain the phenomenon of a classical world as well as the apparent occurrence of quantum jumps or stochastic "events" – see Sect. 4 or (Zeh, 2013) for a historical overview of the conceptual development of quantum theory. So I had never felt any motivation to think of "hidden variables" or any other physics *behind* the successful wave function.

Therefore, I was very surprised on my arrival in Varenna to hear everybody discuss Bell's inequality. It had been published a few years before the conference, but I had either not noticed it or not regarded it as particularly remarkable until then. As this inequality demonstrates that the predictions of quantum theory would require any *conceivable* reality possibly underlying the nonlocal wave function to be nonlocal itself, I simply found my conviction that the latter correctly describes individual reality confirmed. For example, I had often discussed the conservation of total spin or angular momentum in an individual decay process, where it requires nonlocal entanglement between the fragments at any distance in a form that was later called a "Bell state". Therefore, this entanglement cannot represent "just information"; information must be physical – anything else would be homeopathy.

Although the first results from crucial Bell experiments (presented at Varenna by Clauser, Horne, Shimony and others) were still preliminary, they assured me that everybody would share my conviction as soon as Bell's argument had become generally known and understood. I certainly did not expect that fifty years later many physicists would still search for loopholes in the experiments or for justifications of non-locality beyond the wave function, or even deny any microscopic reality in order to avoid contradictions or absurd consequences that result from the prejudice of a reality that has to be local (such as in terms of particles or fields).

So I was particularly glad to hear about John's announcement of a talk "On the assumption that the Schrödinger equation is exact" one or two years after Varenna at a meeting Bernard d'Espagnat had organized in Paris. This title seemed to represent my own ideas, but I will have to come back in Sect. 2 to what he really meant.

Before he published his inequality in 1964, John Bell had shown von Neumann's refutation of hidden variables to be insufficient. (The publication of this paper had been delayed until 1966 by some accidents.) Von Neumann's claim had often been cited in order to defend the Copenhagen interpretation with its irrational "complementarity" concept against such proposals. In Varenna, Bell began his talk by arguing that all physical systems are described by means of two different concepts: classical properties $\Lambda$ and a wave function $\psi$. The latter he suspected to be merely "subjective" (the traditional argument for searching for hidden variables). Today we would then call it an epistemic concept, representing incomplete information, but "information" would only make sense for him with respect to the essential questions "about what?" and "by whom?" This remains true although an objective "state" of incomplete information (an unspecified en-semble of unknown elementary states) may then be *operationally* defined by a certain (incomplete) preparation procedure.

It was this kind of clarity in pointing out misconceptions that always impressed me in discussions with John, or in his talks and publications (Bell, 1987). He never shared the "pragmatic logic" of many physicists who regard arguments as correct just because they somehow lead to the expected or empirically known result. Another example was his objection to some operational arguments used at the conference by axiomatic quantum theorists who suggested the application of certain "superselection rules" in order to replace superpositions by ensembles whenever the required observables were not realizable for some general reason. He insisted that not being able in practice to distinguish between a superposition and an ensemble consisting of its components does not prove them to be the same. This conceptual confusion may also occur in connection with decoherence when one uncritically interprets the reduced density matrix of a subsystem as representing an ensemble rather than entanglement (see Sect. 3). Bernard d'Espagnat had already coined the terms "proper" and "improper mixtures", respectively, to distinguish these two cases. A related third example that comes to my mind is his very politely formulated criticism of Hepp's attempt to justify ensembles of measurement outcomes by means of the purely formal limit of an infinite number of subsystems or degrees of freedom (Bell, 1987: Ch. 6).

The assumption of two different realms of physics (quantum and classical: $\psi$ and $\Lambda$) represented consensus among most physicists at that time – even though one knew from the early Bohr-Einstein debate that macroscopic variables, too, had to obey the uncertainty principle in order to avoid inconsistencies. However, in contrast to the majority of physicists, most participants at the conference agreed that the absence of a well-defined borderline between these two realms represented a severe defect that seemed to call for new physics. Decoherence was not yet known as a possible *effective* borderline, while mesoscopic quantum physics had hardly been seriously studied yet. In fact, when I began presenting decoherence arguments to my colleagues in those years, the usual objection was that "quantum mechanics does not apply to the environment".

John then continued his talk by explaining his arguments against von Neumann's exclusion of hidden variables, gave an outline of David Bohm's theory (which had motivated these arguments), and finally derived his inequality, whose violation, predicted by quantum theory, would exclude *local* hidden variables if confirmed by experiment. This conclusion seemed to be a great surprise and to appear almost unacceptable to many participants. Some young and also some not-so-young physicists there were strongly motivated by dialectic materialism (this conference took place in Italy, two years after 1968!). They could not accept any "idealistic" interpretation of physical phenomena, and sometimes tried to propose very naïve classical models that had all to be in conflict with quantum theory somewhere. However, Bohr had correctly concluded already in 1924 (when his attempt with Kramers and Slater had failed) that "there can be no simple solution" to the problems presented by the quantum phenomena. Nonetheless, in Bell's (and my) opinion this was no reason to abandon the whole concept of reality, which in a theory must be reflected by a consistent, universally valid and successful description of Nature. For him, the renunciation of reality would be the end of physics (as I understood him). Very probably, this conviction was the major motivation in all his endeavors regarding the foundation of quantum mechanics, but his theorem revealed that quantum reality must be in strong contrast to traditional concepts.

At Varenna, I was particularly interested in Bryce DeWitt's talk on the Many Worlds interpretation, because I had mentioned Everett's ideas myself as the only remaining (but possible) solution if the Schrödinger equation was assumed to be exact, universal and complete. But I felt a bit confused when I saw DeWitt translate Everett into the Heisenberg picture. For me, Everett's main point was a unitarily evolving wave function of the universe. He had attended lectures given in Princeton by von Neumann, who had described the measurement process in purely wave mechanical terms, assuming the pointer position to be represented by a moving narrow wave packet rather than a classical variable. This picture of quantum mechanics (which Wigner always meant when he spoke of its "orthodox interpretation") seems to have also influenced Richard Feynman (Zeh, 2011). Only much later did I understand that for DeWitt and David Deutsch, "Many Worlds" meant many trajectories in configuration space (or many Feynman paths), while for Everett and me this concept meant many, in excellent approximation dynamically "autonomous", wave packets, which may possibly even form an overcomplete set (see Sect. 4). For example, while Deutsch regards a quantum computer as an example for many worlds in action, in Everett's sense they must all remain part of one "world" in order to lead to one quasi-classical result that can be observed and used by humans. Only if there were *macroscopically* different intermediary states of the com-puter, could their superposition give rise to different "worlds" by their decoherence – and thus ruin the quantum computer. These different formal representations of "reality" (by classical configurations or by wave packets) are also relevant to Bell's different interpretations of quantum mechanics, which I will now discuss.

**2. Bell on Bohm's Theory**

Although Bohm and Hiley were present at Varenna (as well as Andrade e Silva, who represented Louis deBroglie), I first understood Bohm's theory (Bohm, 1952) when studying Bell's Varenna contribution (Bell, 1987: Ch. 4). He presented this theory as a "simple example" for hidden variables, even though it was in contrast to his introductory remarks: it neither *replaced* the wave function $\psi$ nor explained it in terms of an ensemble of hidden variables. In more recent language, this theory is "$\psi$-ontic", but in addition it assumes the existence of hidden variables $\lambda$ that are here identified with the pre-quantum variables (such as particle positions and field amplitudes): it is not "$\psi$-complete". So these variables are isomorphic to the arguments of his wave function, while the appearance of particles (such as photons) for all quantum *fields* remains an open problem. Nonetheless, this theory allowed Bohm to assume the Schrödinger equation to be exact (the same as in Everett's later theory!), and a classical configuration of the world to be dynamically guided by the arising wave function instead of obeying Hamilton's equations. Bell meant essentially Bohm's theory by the title of his talk that I first heard in Paris, where the Schrödinger equation is assumed not only to be exact, but also to be universal. There are no strictly classical variables $\Lambda$ any more (they are simply functions of the $\lambda$'s), but Bell regarded it as an important advantage that Bohm's theory does not need the "notoriously vague concept of a reduction of the wave packet".

However, he also remarked that "what happens to the hidden variables during and after a measurement is a delicate matter". In my opinion this is a serious weak point of the theory, since the $\lambda$'s have to be *postulated* to form a statistical distribution with probabilities given by $|\psi(\lambda)|^2$, while only one of the trajectories is assumed to describe reality. Although this distribution is dynamically consistent with Bohm's dynamics, (1) no plausible motivation for this statistical assumption (in contrast to the individually treated wave function) is given, and (2) the probabilities would have to change under a change of information by measurements, although no physical carrier of this information is taken into account ("information by whom?"). This comes close to the crucial assumption of an external (human?) observer in the Copenhagen interpretation.

Supporters of Bohm's theory often present special applications in order to illustrate it, although it is evident already from its general construction that *all* its observable predictions must be in accord with traditional quantum mechanics. Therefore, they cannot serve to confirm Bohm's theory. However, these trajectories appear plausible only in simple cases, such as single-particle scatterings. In general, they may have very surprising properties, and little to do with what one would expect or what we *seem* to observe (Zeh, 1999). In my opinion, this fact eliminates the major motivation for this theory, since its "traditionalistic" trajectories can neither be observed nor remembered: they are observationally meaningless.

On the other hand, Bohm was perhaps the first physicist to seriously consider entangled wave functions for macroscopic systems. Shelly Goldstein even claimed that Bohm anticipated the decoherence concept when he discussed measurements in his theory. This is a bit of an overstatement and a misunderstanding. In order to describe *successions* of measurements, Bohm had to discuss how the probability distribution of his classical configurations $\lambda$ is restricted by all previous measurements to the carrier of some small "effective" component of the wave function (essentially identical with "our" Everett branch), and this means first of all that these branches have to remain dynamically autonomous for some time (the way we are using wave functions in practice). This is similar to Mott's early treatment of $\alpha$-particle tracks in the Wilson chamber, which did *not* yet take into account subsequent decoherence of the droplet positions by their entanglement with an unbounded environment. Only this *real* (irreversible) decoherence explains why different "quasi-classical" wave packets forming one superposition never meet again in configuration space in order to interfere, that is, why the required autonomy holds "forever" in practice. Within these autonomous branches, wave functions for macroscopic variables are restricted to narrow wave packets that resemble classical points. Bohm might then have noticed that his presumed fundamental variables $\lambda$ would become obsolete if these branches themselves were "selected" in some sense. In mesoscopic cases, decoherence theory requires detailed calculations for realistic environments, which were performed during the eighties by Wojciech Zurek, Erich Joos, and many others.

During the decade following Varenna, John Bell presented various versions of his talk about the "assumption that the Schrödinger equation is exact". Like many other fundamental papers at that time, they were often first published in the informal Epistemological Letters, since established journals were still reluctant to accept papers on interpretational issues of quantum theory. Only after his inequality had become known to allow crucial experiments to be performed in laboratories, did this situation slowly change – one of John's historically most important achievements.

A slightly modified and extended version of these talks (for a special purpose) was published in 1981 under the new title "Quantum mechanics for cosmologists" (Bell 1987: Ch. 15). It contains a number of important statements. Talking about Bohm, he says that "nobody can understand this theory until he is willing to think of $\psi$ as a real objective field rather than a probability amplitude". This is in explicit contrast to his introductory remarks at Varenna about $\psi$ as a "subjective" concept. As only one set of $\lambda$'s is assumed to be real (representing one point somewhere in the myriads of branches of the universal wave function), he compares $\psi$ with the Maxwell fields, which are similarly assumed to exist even where no charged "test particles" are present. But he adds that "it is in terms of the $\lambda$" (which he now calls $x$) "that we would define a psycho-physical parallelism – if we were pressed to go so far". Therefore, he now called Bohm's "hidden" variables "exposed", although their exposure (together with their very existence) remains a model-specific hypothesis. The $\lambda$'s may appear "more real" than $\psi$ to the traditional mind because they are defined as *local* "beables". This is also why observable quantum non-locality is often understood as requiring a spooky action at a distance rather than the consequence of a nonlocal beable: the "real" wave function. (In classical context, we similarly prefer to believe that we see objects rather than – more realistically – the light reflected by them, or even the nerve cells excited by the light in the retina and in the brain. In this classical picture, however, all these physical elements and their interactions are local and can be regarded as empirically well established.)

When mentioning Everett's interpretation as another possibility for the Schrödinger equation to be exact, John usually disregarded it as "extravagant" – not for being wrong (Bell 1987: Ch. 20). This position appears natural from a traditional point of view. Similarly, Stephen Weinberg declared in an interview about his recent book on quantum mechanics (for *Physics Today Online* of July 2013) that "this effort [of not conceptually distinguishing the apparatus or the physicist from the rest of the world] may lead to something like a 'many worlds' interpretation, which I find repellent." But he had to add: "I work on the interpretation of quantum mechanics from time to time, but have gotten nowhere." In fact, there are strong emotions but hardly any convincing arguments against Everett. This has even led to some "mobbing" by traditionalists of all kinds against Everettians or even against a fundamental role of decoherence. In (Bell, 1987: Ch. 11), Bell raised the objection that Everett's branches are insufficiently defined or arbitrary, but precisely this ambiguity has been removed by decoherence (see Sects. 3 and 4).

After John had given a version of his talk at Heidelberg in about 1980, we had a brief correspondence, where I tried to point out to him that Bohm's theory is just as extravagant as Everett's in the sense that Bohm's wave function contains the same myriads of components that are regarded as "many worlds" by Everettians. The only difference is that all but one of them are called "empty" by Bohmians. Nonetheless, all the empty parts of the wave function are assumed to *exist*, too, in order to avoid the collapse! We also debated the relation between the concept of reality and that of "heuristic fictions" in physics on this occasion, but the correspondence led to no obvious result. However, it may have had some consequences a few years later (see Sect. 3).

When re-reading Bell's "Quantum theory for cosmologists" for the preparation of this paper, I discovered another astonishing remark about Everett. Bell initially points out that he is not quite sure whether he understands Everett correctly, but then claims a "previously unknown close relationship between Everett and Bohm". He says (surprisingly) that "all instantaneous classical configurations $\lambda$ are supposed to exist" in Everett's theory (the assumption that he regarded as extravagant). This interpretation may come close to Deutsch's identification of (many) "worlds" with a continuum of trajectories in configuration space (cf. Sect. 1). Deutsch has indeed repeatedly called Bohm's theory a "many-worlds theory under permanent denial". Bell's remark indicates that he, too, would prefer *beables* to be local – probably a major reason for his favor for Bohm's theory. So Bohm and Deutsch *presume* classical concepts (points in configuration space); this explains why they both do not need decoherence to justify them.[*] If we did instead define "worlds" to consist of trajectories for macroscopic objects plus wave functions for microscopic ones, we would be back searching for Bohr's borderline between two different realms of physics, now using much improved but nonetheless as yet unsuccessful experimental techniques. In contrast, Everett interpreted the world completely and solely in terms of wave functions (he was von Neumann's student). A relation to classical concepts may then be provided only in terms of wave packets in configuration space. This means that Everett is conceptually *not* closely related to Bohmian mechanics with its classical variables $\lambda$, but rather to Bell's favorite-to-come: collapse theories.

---

[*] Note added after publication: This may similarly apply to Vaidman's definition of branches in Sect. 12.5 of this book if he requires them not to contain any nonlocal entanglement (such as total spin states for separate particles). Such states do not require any action at a distance in order to violate Bell's inequality, but they are nonetheless nonlocal.

**3. Collapse Theories**

In 1987, John Bell surprised his admirers by a drastic change of mind. Inspired by (Ghirardi, Rimini and Weber, 1986), he now advocated for collapse theories (Bell, 1987: Ch. 22). That is, he dropped the assumption that the Schrödinger equation is exact (as it is in Bohm's mechanics), and instead supported what he had previously called the "notoriously vague collapse" – albeit in a newly specified, hypothetical form. If correct, this proposal would avoid all those myriads of "other" branches of the wave function which he found extravagant in Everett's interpretation, and which had to be regarded as "empty" in Bohm's. This does not necessarily mean that he abandoned Bohm completely. He may simply have started an independent attempt to search for a solution of the quantum problems in terms of a realistic theory, but his radical change of concepts may also indicate that he was not quite happy any more with his previous favorite.

When Johann von Neumann first formulated his collapse or reduction of the wave function in Chaps. V and VI of his book (von Neumann, 1932), he felt motivated by the need not only to explain definite pointer positions, but also to facilitate a psycho-physical parallelism that is applicable to local observers in spite of the nonlocality of the generic wave function. These two different though related intentions reflect Bohr's and Heisenberg's slightly different understandings of quantum measurements. While the former insisted that indeterministic measurement outcomes have to be objectively described in terms of classical pointer states (which could thereafter be observed in a traditional way by interaction with classical media and observers), the latter had originally regarded the measured properties (including particle positions) as being *created by their observation by humans*. This difference left many traces in the history of quantum measurement theory, but both aspects seem to be relevant in some way even for an ontic interpretation of the wave function (see Sect. 4).

The GRW collapse was clearly meant to describe an objective physical process (for a phenomenon that Bohr had regarded as *not* dynamically analyzable). Therefore, these authors concentrated on a process of "spontaneous localization" for the wave functions of macroscopic variables. For this purpose, they postulated a non-unitary and irreversible "master equation" instead of the von-Neumann equation for the density matrices of all isolated physical systems. The precise form and strength of its non-unitarity had to be adjusted in order to describe Born's probabilities as part of this new fundamental dynamical law. Von Neumann's equation is equivalent to the linear Schrödinger equation, which they now assumed to apply only approximately in a microscopic limit. They also assumed tacitly that this density matrix describes an (ever-growing) ensemble of *potential* wave functions, but the problem is that such an ensemble is not uniquely determined by the density matrix, nor can the latter distinguish between ensembles and an entanglement of the considered system with others.

Indeed, immediately after their paper had appeared, Erich Joos was able to demonstrate (Joos, 1987), that their master equation can be well understood, and even be made precise, within unitary quantum mechanics as a consequence of the unavoidable interaction of macroscopic systems with their realistic environment (later called decoherence). This mechanism could be quantitatively confirmed in several mesoscopic cases, while it obviously applies to all macroscopic ones. However, it describes growing entanglement rather than a transition from pure states into ensembles (such as those of different measurement outcomes). Therefore, two questions arise: (1) how would GRW's master equation have to be interpreted in order to describe measurements, and (2) what does the undeniable environmental decoherence, which can hardly *accidentally* lead precisely to the required density matrix (*as though* it were an ensemble), mean for the measurement process?

In order to answer the first question, John Bell proposed a nonlinear stochastic ("quantum Langevin") equation for the dynamics of individual wave functions. This new equation would thus have to replace the Schrödinger equation. The ensemble of potential future wave functions thereby arising can be represented by a density matrix that would then obey GRW's master equation. His specific model postulated that jumps of single-particle wave functions into slightly more localized partial waves occur spontaneously with Born-type probabilities. He assumed the time scale for these jumps to be of order $10^{15}$ s, but for the center of mass of a multi-particle object this time scale would have to be divided by the number of particles, and so become sufficiently short for such collective variables. However, he also noticed and listed a number of problems, such as the entanglement between particles or the generalization of his proposal to QFT (others have later been added), although he expressed hope that they can be overcome. I doubt that this has ever been achieved for this kind of model, but there exists a wealth of similar and also quite different possibilities for a collapse mechanism, which can be falsified only one after another and when defined exactly. Only in that case would they share this property of being falsifiable with Everett's interpretation, which could be ruled out by the discovery of an appropriate violation of global unitarity (Arndt and Hornberger, 2014). Therefore, the *possibility* of a dynamical collapse still exists in principle, and in spite of the fact that none of its proposed versions has ever been verified by experiments. While the confirmation of such hypothetical non-unitarity *would* close this fundamental debate forever, environmental decoherence must remain important, as it seems to describe all as yet observed *apparent* (local) non-unitarities.

The major reason for this undecided situation is that any conceivable collapse dynamics would have to be carefully shielded against all competing decoherence effects in order to be confirmed – an almost impossible requirement in the macroscopic realm. As the reduced density matrix arising from decoherence cannot be locally distinguished from that of an ensemble, it is sufficient FAPP (for all practical purposes – a later often misused term that Bell invented in his last paper "Against measurement" (Bell, 1989)).

However, interaction with the environment can never describe the transition of a global pure state into an ensemble of possible outcomes – it merely describes the dislocalization of all macroscopic superpositions by means of their spatially spreading entanglement. Even if the environment or the apparatus were described by an *ensemble of different initial states* (incomplete information about microscopic initial conditions), as often suggested in the hope that this ensemble might then lead to the expected ensemble of different outcomes, the conclusion of a resulting *superposition* of different outcomes would remain valid for each of its individual members. This very general argument has often been emphasized by Eugene Wigner, who pointed out that the density matrix characterizing the initial ensemble merely *hides* the lasting entanglement. The latter must ultimately give rise to a global superposition of "many worlds" for each *individual* state. For similar reasons, no kind of classical "noise" (represented by an uncertain Hamiltonian, for example) would explain the required ensemble, whereas the often mentioned interaction with quantized gravity is quantum mechanically just a special (though not very relevant) form of environmental decoherence. A genuine collapse would have to be explicitly postulated and empirically confirmed as a fundamental deviation from unitarity. The omni-present formation and spreading of initially absent entanglement, on the other hand, seems to form the general "master arrow of time" characterizing our universe. In this way, it also forms the physical basis for the general concept of time-asymmetric "causality" (Zeh, 2007).

In contrast to Bohr's above-mentioned understanding, collapse theories assume the wave function (though not the Schrödinger equation) to apply universally, and therefore to form a complete ontic concept. This wave function must then in principle also describe the brain with its expected specific role in a psycho-physical parallelism. In the absence of Bohm's $\lambda$'s, and under his new assumption of spontaneous jumps, Bell now suggested that consciousness be related to such (again model-specific) "events", while von Neumann had assumed consciousness to be related in the sense of a psycho-physical parallelism to the observers' quantum *states*, which he assumed to be formed by means of his vaguely defined collapse. Some kind of a psycho-physical connection is certainly required in order to understand how our subjective observations are related to the hypothetical real world (that is, how subjective observations come about in objective terms). Einstein spoke of the "whole long chain of interactions from the object to the observer" that we must understand in order to know *what* we have observed.

## 4. Consequences of decoherence

Already in my first paper on decoherence (Zeh, 1970), I had explicitly pointed out that entanglement with the environment (in that paper called "quantum correlations") does *not* lead to the apparently required ensemble of possible outcomes (a proper mixture). It merely explains the absence of certain *local* superpositions, which had often been attributed to fundamental "superselection rules" or a global stochastic collapse. Since no hints of a collapse had ever been directly observed, I suggested an interpretation similar to Everett's – see also (Zeh, 2013: Sect. 4) for details. (Some Oxford philosophers have recently "rediscovered" this successful combination of decoherence and Everett.) This very possibility is sufficient to demonstrate that Bell's claim that "the wave function must either be incomplete or not always right" cannot be upheld. In several subsequent papers during the seventies I even tried to learn more about the physical localization of consciousness in quantum mechanical terms by using the single-sum Schmidt canonical representation for entangled states (assuming a fundamental local observer system in the brain, for example), but this attempt did not turn out to be particularly helpful. Objectively, we can argue only in terms of "robust" properties, such as memory (physically realized in data storage devices or by decohered variables in the brain – even though these are only intermediary concepts in the long chain between object and subject). Such robust states describe quasi-classical properties, which may form discrete sets for digital devices or neuronal frameworks. In the general case, they were later called "pointer states" by Zurek (Zurek, 1981), whose readers, however, may have misunderstood them according to the title of this paper as forming a proper mixtures after a decoherence process – see also (Camilleri, 2009). Precisely this "naïve" misunderstanding of decoherence seems to have considerably contributed to its popularity, while the introduction of new names, such as "einselection", "quantum Darwinism" or "existential interpretation", did not add any new contents to this theory.

Most workers in the field of decoherence have indeed restricted their interest to the effects of the environment on the density matrices of local systems. This is sufficient for *most* practical purposes, and it allows one to understand the pseudo-concept of "complementarity" simply as a consequence of different couplings to the environment by means of different measurement devices. Because of this decoherence phenomenon, existing proposals for a fundamental collapse are mostly attempts to mimic the effects of environmental decoherence. They are thus based on a prejudice that arose before decoherence was understood as a unitary process. However, there are far more *other* possibilities for a collapse along Einstein's "whole long chain" which could finally explain individual states of awareness (all we know with certainty) in an objective dynamical way. As decoherence describes *apparent* transitions into ensembles and *apparent* quantum jumps, which can even be experimentally confirmed as forming smooth processes, it has indeed often been misinterpreted as a derivation of the probabilistic collapse from the Schrödinger equation. (I remember authors claiming that decoherence saves us from the conclusion of Many Worlds, although precisely the opposite is true!) This misuse of the density matrix has a long tradition in measurement theory, and John was certainly right to object against it also in connection with decoherence. But his critical position is then often misused in turn as an argument to entirely dismiss the essential role of decoherence in measurements. In order to understand what decoherence "really" means, one has to analyze the consequences of interactions with the environment on the *individual* global wave function (similar to Bell's formulation of the stochastic collapse mechanism for individual states described in Sect. 3).

The essential insight that originally led to the concept of decoherence was that, for dynamical reasons, entanglement must be far more common in macroscopic systems than had ever been envisioned or taken into account. It had been overlooked as long as unitarity was believed to apply only to microscopic systems. Ironically, though, global unitarity is able to explain local non-unitarity. However, it is *not* the formal diagonalization of the thereby arising reduced density matrices of all systems that explains "effective ensembles", but rather the fact that an ever-increasing number of components of the wave function become dynamically independent of one another ("autonomous") in this way. In particular, this progressive entanglement means that nonlocal superpositions cannot, in practice, be relocalized any more in order to describe some "recoherence". This situation is analogous to the justification of Boltzmann's *H*-theorem in deterministic particle mechanics, where the $\mu$-space distribution is very unlikely to be later affected within reasonable times by nonlocal many-particle correlations that were previously created in chaotic collisions. These correlations must nonetheless persist in order to preserve Gibbs's ensemble entropy for distributions on $\Gamma$-space. Similarly, the permanent "branching" of the wave function into autonomous components that can never interfere any longer is quite compatible with Schrödinger's determinism and the high dimensionality of configuration space (using plausible cosmic initial conditions). As long as there are no empirical indications for a collapse mechanism, the existence of unobservable branches of the wave function ("worlds") in addition to the observed one is thus not a matter of "philosophy", but of dynamical consistency.

One may then easily recognize that different pointer positions, dead and alive cats, *and* different states of awareness of an observer (but *none* of their superpositions) can only exist separately within such autonomous branches of the wave function. Therefore, the various autonomous "versions" of each observer that similarly arise from decoherence can only represent *separate* "subjective identities" ("many minds") – even though they may have shared their *early* histories. Because of their emerging dynamical autonomy, they can only remember the histories of their own branches, including consequences of superpositions that had existed *before* their decoherence. Statistical weights according to the squared norms of the branch wave functions have to be postulated for the sole purpose of correctly "selecting" *our subjectively* observed version by chance; they have no meaning from the objective "birds perspective" that is described by unitary dynamics. These empirically observed weights, which explain the frequencies of results in series of measurements *observed by us* ("Born's rule"), cannot be derived from the objective part of the theory, but they are the only dynamically consistent ones in the sense that they are conserved under the Schrödinger dynamics. In contrast to other conceivable weights they are thus not changed by later branchings – and so give rise to the concept of "consistent histories" in terms of effective wave functions.

So we have to conclude that the indeterminism we observe in the quantum phenomena does not reflect an objective dynamical law. Rather, it is a consequence of the branching histories of all observers in a quantum world. While, in the classical description, the subjective decision about who is "you" among all conscious observers existing in the deterministically evolving universe occurs only once for your lifetime, it is repeated many times every second with respect to your permanently arising new "versions". At least, this would be the consequence of global unitarity; it is not just a fiction. It is also the reason that a wave function often appears to us as representing "just information". One may regard a theory as incomplete if it does not determine the individual subjective observer, but this "defect" would then also apply to classical (Laplacean) theories.

Slightly generalizing John Bell's terminology, one may say that the autonomy of individual branches arising from decoherence allows us to replace the formal "plus" of their superposition by an effective phase-independent "and", while an "or" is meaningful only with respect to the individual version of a potential observers. In this Everettian sense, Heisenberg's subjective interpretation of measurement outcomes as being *created* by their observation may thus be justified at last – although not in terms of fundamental particle or other classical concepts. Without taking into account this "subjective individualization" of measurement results (an effective projection in Hilbert space) we could not even prepare pure states for microscopic systems in the laboratory, where we perform series of measurements for this purpose in order to continue the experiment only with the appropriate outcomes.

Unfortunately, John Bell never seriously considered this version or variant of Everett (as far as I know). He may still have regarded it as "extravagant" because of the myriad versions of each observer that have to arise according to the Schrödinger equation. Collapse models, in contrast, may *postulate* that similarly defined branches of the wave function all but one disappear from reality, but an observer, who can only exist in a branch, does not have to bother whether many other versions of himself do exist in other autonomous branches, or rather have disappeared from reality – unless he is one of those rare non-pragmatic physicists such as John Bell.

A pragmatic physicist will in any case *use* the collapse FAPP as soon as decoherence (understood as the dislocalization of initially microscopic superpositions) has become "real" rather than virtual (irreversible in practice) after a measurement-like process, in which a microscopic difference led to different macroscopic consequences. This onset of environmental decoherence defines a natural position for the Heisenberg split. Decoherence also allows us approximately to describe all robust properties of macroscopic systems in classical terms, while most others can locally be treated by conventional statistical methods, such as retarded master equations (Zeh, 2007).

The popular objection that "this apparent collapse applies only FAPP" means no more than that this solution of the problem is not what John Bell (and many other physicists) had expected and hoped for. It is nonetheless sufficient, without being based on any novel and speculative postulates. We have learned from Everett that we do not have to expect the collapse to represent an objective physical process that may some day be further specified and confirmed by experiments (although the possibility of such a genuine non-unitarity can never be excluded with certainty). The "effective collapse FAPP" is just a convenient picture, representing the situation of the subjective observer in his changing branch. So it cannot (and need not) be defined *exactly*, and it may even be assumed to "act" superluminally. Events may consistently be assumed to "occur" even in cases where they will never be observed: the phase relations which must still exist between different "world" components according to the Schrödinger equation become irrelevant for all *potential* observers in their branches as a consequence of decoherence. However, this reasonable convention (in a dynamically consistent quantum world that may be assumed to describe reality) seems to be the source of many misconceptions, such as counterfactuals, new logics, complementarity, an "uncertain" reality, "it from bit", and similar terminology that John Bell found neither very helpful nor meaningful.

**Acknowledgement:** I wish to thank Shelly Goldstein, Basil Hiley, Erich Joos, and Claus Kiefer for their criticism, comments, and discussions regarding earlier versions of the manuscript.